\documentclass[twocolumn,prd]{revtex4}
\usepackage{amssymb}

\usepackage[english]{babel}

\usepackage{amsmath,amssymb,amsxtra,amsfonts}
\usepackage{graphicx}
\usepackage{multirow}
\usepackage{dcolumn}
\usepackage{txfonts}
\def\bea{\begin{eqnarray}}
\def\eea{\end{eqnarray}}
\def\be{\begin{equation}}
\def\ee{\end{equation}}

\begin{document}

\title{Constraints on Cosmic Distance Duality Relation from Cosmological Observations}

\author{Meng-Zhen Lv$^{a}$}
\author{Jun-Qing Xia$^{b}$}\email{xiajq@bnu.edu.cn}
\affiliation{$^a$Key Laboratory of Particle Astrophysics, Institute of High Energy Physics,
Chinese Academy of Science, P. O. Box 918-3, Beijing 100049, China}
\affiliation{$^b$Department of Astronomy, Beijing Normal University, Beijing 100875, China}

\begin{abstract}

In this paper, we use the model dependent method to revisit the constraint on the well-known cosmic distance duality relation (CDDR). By using the latest SNIa samples, such as Union2.1, JLA and SNLS, we find that the SNIa data alone can not constrain the cosmic opacity parameter $\varepsilon$, which denotes the deviation from the CDDR, $d_{\rm L} = d_{\rm A}(1+z)^{2+\varepsilon}$, very well. The constraining power on $\varepsilon$ from the luminosity distance indicator provided by SNIa and GRB is hardly to be improved at present. When we include other cosmological observations, such as the measurements of Hubble parameter, the baryon acoustic oscillations and the distance information from cosmic microwave background, we obtain the tightest constraint on the cosmic opacity parameter $\varepsilon$, namely the 68\% C.L. limit: $\varepsilon=0.023\pm0.018$. Furthermore, we also consider the evolution of $\varepsilon$ as a function of $z$ using two methods, the parametrization and the principle component analysis, and do not find the evidence for the deviation from zero. Finally, we simulate the future SNIa and Hubble measurements and find the mock data could give very tight constraint on the cosmic opacity $\varepsilon$ and verify the CDDR at high significance.

\end{abstract}

\maketitle

\section{Introduction}
\label{introduction}

In 1998 the analyses of the redshift-distance relation of type Ia supernova (SNIa) at low redshift $z < 2$ have demonstrated that the Universe is now undergoing an accelerated phase of expansion \cite{riess1998supernova,perlmutter1999measurements}. Currently, cosmological observations have provided tight constraints on distance measures: the luminosity distance $d_{\rm L}$ by measuring the SNIa and the angular diameter distance $d_{\rm A}$ by measuring the baryon acoustic oscillations (BAO), which can be used to constrain different cosmological parameters in various theoretical models \cite{planck2015fit}. In general, the luminosity distance and the angular diameter distance should satisfy the well-known cosmic distance duality relation (CDDR):
\begin{equation}
d_{\rm L} = d_{\rm A}(1+z)^2~.\label{cddr}
\end{equation}
This relation, which is also called ``Etherington relation'' in the literature, holds only for the validity of three fundamental conditions:
\begin{itemize}
\item The spacetime is described by a metric theory of gravity;
\item Photons travel along unique null geodesics;
\item The number of photons is conserved.
\end{itemize}
Therefore, any departure from these three conditions, such as the deviation from a metric theory of gravity, photons not traveling along null geodesics and the variation of photon number, will reveal the new physics beyond the standard model.

In the literature, in order to test the CDDR, a model-independent method has been widely used in which people use the current datasets of $d_{\rm L}$ from SNIa or Gamma-Ray Bursts (GRB) measurements and $d_{\rm A}$ from BAO or X-ray measurements at the same redshift to constrain the parameter $\eta=d_{\rm L}/d_{\rm A}(1+z)^2$ (e.g. see refs. \cite{2010ApJ...722L.233H,2011arXiv1104.2497L,2012ApJ...745...98M,2011RAA....11.1199C,DeBernardis:2006ii,
2011SCPMA..54.2260C,Goncalves:2011ha,Liao:2012bg,Ellis:2013cu,Bassett:2003vu,Bassett:2003zw,Uzan:2004my,Nair:2011dp} and references therein). If $\eta$ obtained from the $d_{\rm L}$ and $d_{\rm A}$ datasets is different from the unity, the CDDR relation is violated. Recently, Ref.\cite{cddrzhu} used a new compilation of strong lensing system to extract the information of $d_{\rm A}$ and obtained the constraint on the parameter: $\eta=-0.004^{+0.322}_{-0.210}$ ($68\%$ C.L.), together with the ``Joint Luminosity Analyses'' (JLA) compilation of SNIa \cite{2014A&A...568A..22B}. Apparently, this method for testing CDDR is conservative and independent on the underlying cosmological model. However, the big problem is that current observations cannot provide the information of luminosity distance $d_{\rm L}$ and angular diameter distance $d_{\rm A}$ for an astronomical target at same time. Therefore, they have to use the information of $d_{\rm A}$ from the galaxy cluster observations and $d_{\rm L}$ from the SNIa measurements at the similar redshift, which inevitably brings large numerical errors on the determination of $\eta$.

On the other hand, there is another model-dependent method to study this relation. The CDDR is in deep connection with the cosmic opacity \cite{Nair:2012dc,More:2008uq}. A variation of photon number during propagation towards us, which could be caused by some simple astrophysical effects, like the interstellar dust, gas and/or plasmas, and some exotic physics beyond the standard model, will affect the SNIa luminosity distance measures but not the determinations of the angular diameter distance in a certain underlying cosmological framework, and consequently modify the CDDR relation. Assuming $\tau(z)$ denotes the cosmic opacity between an observer at $z = 0$ and a source at $z$, the flux received from the source would be attenuated by a factor $e^{-\tau(z)}$. Then the luminosity distance has
\begin{equation}
d_{\rm L,obs}(z) = d_{\rm L,true}(z)\exp(\tau(z)/2)~,
\end{equation}
because intensity is inversely proportional to square of distance between the source and the observer. Ref. \cite{cddrpara} introduced a parameter $\varepsilon$ to study deviations from the Etherington relation of the form
\begin{equation}
d_{\rm L}(z) = d_{\rm A}(z)(1+z)^{2+\varepsilon}~,
\end{equation}
where $\varepsilon$ denotes the departure from the transparency. Considering the small value of $\varepsilon$ at low redshift, this is equivalent to assume an optical depth parameterization $\tau(z)=2\varepsilon z$. The advantage of this method is that we can use the measurements of $d_{\rm L}$ with high precision to constrain the cosmic opacity and avoid to include the large uncertainties from the measurements of galaxy clusters. Currently, the most tight constraint comes from the dataset combination of ``Union2 Compilation'' SNIa sample and the Hubble parameter as a function of redshift $H(z)$: $\varepsilon=-0.01^{+0.08}_{-0.09}$ ($95\%$ C.L.) \cite{2010JCAP...10..024A}. Until now, all the measurements satisfy the CDDR relation at $68\%$ confidence level.

In this paper we mainly focus on the model-dependent method to verify the CDDR relation and update the constraints on the cosmic opacity from the latest measurements on SNIa samples of ``Union2.1 Compilation'' (Union2.1) \cite{2012ApJ...746...85S}, ``Joint Luminosity Analyses'' (JLA) \cite{2014A&A...568A..22B} and ``Supernovae Legacy Survey'' (SNLS) \cite{2011ApJS..192....1C}, as well as the measurement on the Hubble parameter $H(z)$. Furthermore, we also include the GRB, BAO and the distance information of cosmic microwave background (CMB) into the analyses to help improving the constraints on cosmic opacity. The paper is organized as follows: In Section~\ref{data}, we introduce the datasets used in the analyses. We present the numerical results in Section~\ref{results}. Finally, in Section \ref{summary} conclusion and discussion are drawn.

\section{Current Datasets}
\label{data}

In our calculations, we rely on the following current observational data sets: i) SNIa and GRB distance moduli; ii) Hubble parameter determinations; iii) BAO in the galaxy power spectra; iv) CMB distance information.

\subsection{Type-Ia Supernovae \& Gamma-Ray Bursts}

The SNIa distance moduli provide the luminosity distance as a function of redshift $z$. In this paper, we use the latest SNIa Union2.1 compilation of 580 dataset from the Hubble Space Telescope Supernova Cosmology Project \cite{2012ApJ...746...85S}. The data are usually presented as tabulated distance modulus with errors. In this catalog, the redshift spans $0<z<1.414$, and about 95\% samples are in the low redshift region $z<1$. The authors also provided the covariance matrix of data with and without systematic errors. In order to be conservative, we include systematic errors in our calculations.

For comparison, we also consider the following two SNIa data:
1) 472 samples from the first three year of SuperNova Legacy Survey (SNLS) Program (123 low-z, 93 SDSS, 242 SNLS, and 14 Hubble Space Telescope) at $0.01 < z < 1.4$ \cite{2011ApJS..192....1C};
2) 740 samples from the SDSS-II/SNLS3 Joint Light-curve Analysis (JLA) at redshift up to 1.30 including several low-redshift samples ($z < 0.1$), all three seasons from the SDSS-II ($0.05 < z < 0.4$), three years from SNLS ($0.2 < z < 1$) \cite{2014A&A...568A..22B}.

These two data compilations are different from the Union2.1 in three major aspects:
1) The two supernova nuisance parameters $\alpha$ and $\beta$ coming from light-curve calibration and are handled correctly instead of held at their best fit values; 2) They offer covariance between the light-curve fit; 3) The luminosity distance takes into account the difference between the CMB frame and heliocentric frame redshifts, which is important for some of the nearby supernova. Furthermore, the JLA compilation includes intrinsic dispersion and gravitational lensing effect in supernova magnitude, while the SNLS does not.

In addition, we also consider another luminosity distance indicator provided by GRBs, that can potentially be
used to measure the luminosity distance out to higher redshift than SNIa. GRBs are not standard candles since
their isotropic equivalent energetics and luminosities span 3-4 orders of magnitude. However, similar to SNIa it has been proposed to use correlations between various properties of the prompt emission and also of the afterglow
emission to standardize GRB energetics (e.g. Ref. \cite{2004ApJ...613L..13G}). Recently, several empirical correlations between GRB observables were reported, and these findings have triggered intensive studies on the possibility of using GRBs as cosmological ``standard'' candles. However, due to the lack of low-redshift long GRB data to calibrate these relations, in a cosmology-independent way, the parameters of the reported correlations are given assuming an input cosmology and obviously depend on the same cosmological parameters that we would like to constrain. Thus, applying such relations to constrain cosmological parameters leads to biased results. In Ref. \cite{2008ApJ...680...92L} this ``circular problem'' is naturally eliminated by marginalizing over the free parameters involved in the correlations; in addition, some results show that these correlations do not change significantly for a wide range of cosmological parameters \cite{firmani2007,schaefer2007}. Therefore, in this paper we use the 69 GRBs over a redshift range $z\in[0.17,6.60]$ presented in Ref. \cite{schaefer2007}, but we keep into account in our statistical analysis the issues related to the circular problem that are more extensively discussed in Ref. \cite{2008ApJ...680...92L} and also the fact that all the correlations used to standardize GRBs have scatter and a poorly understood physics.

In the calculation of the likelihood from SNIa and GRBs, we have marginalized over the absolute magnitude $M$ which is a nuisance parameter, as done in Refs.\cite{2001A&A...380....6G,pietro2003}:
\begin{equation}
\bar{\chi}^2 = A - \frac{B^2}{C} + \ln{\left(\frac{C}{2\pi}\right)}~,
\end{equation}
where
\begin{equation}
A=\sum_{i}\frac{(\mu^{\rm data} - \mu^{\rm th})^2}{\sigma_i^2}~~,~~~B=\sum_{i}\frac{\mu^{\rm data} - \mu^{\rm th}}{\sigma_i^2}~~,~~~C=\sum_{i}\frac{1}{\sigma_i^2}~.
\end{equation}

\begin{table*}[t] 
\caption{$H(z)$ measurements and their errors in units of ${\rm km\,s^{-1}\,Mpc^{-1}}$. (${\tt Reference. -}$ [1] Gazta\~naga {\it et al.} (2009); [2] Stern {\it et al.} (2010); [3] Moresco {\it et al.} (2012); [4] Zhang {\it et al.} (2012); [5] Simon {\it et al.} (2005).)}\label{Hz}
\begin{center}
\begin{tabular}{c|c|c|c|c|c|c|c|c|c|c|c|c}

\hline
\hline

$z$ & $~0.07~$ & $~0.09~$ & $~0.12~$ & $~0.17~$ & $0.1791$ & $0.1993$ & $~0.2~~$ & $~0.24~$ & $~0.27~$ & $~0.28~$ & $0.3519$ & $~0.40~$ \\

$H(z)$ & $69$ & $69$ & $68.6$ & $83$ & $75$ & $75$ & $72.9$ & $79.69$ & $77$ & $88.8$ & $83$ & $95$ \\

$\sigma_{H(z)}$ & $19.6$ & $12$ & $26.2$ & $8$ & $4$ & $5$ & $29.6$ & $2.65$ & $14$ & $36.6$ & $14$ & $17$ \\

Ref. & $[4]$ & $[2]$ & $[4]$ & $[2]$ & $[3]$ & $[3]$ & $[4]$ & $[1]$ & $[2]$ & $[4]$ & $[3]$ & $[2]$ \\

\hline\hline

$~0.43~$ & $0.48$ & $0.5929$ & $0.6797$ & $0.7812$ & $0.8754$ & $0.88$ & $0.9$ & $1.037$ & $1.3$ & $1.43$ & $1.53$ & $1.75$\\

$86.45$ & $97$ & $104$ & $92$ & $105$ & $125$ & $90$ & $117$ & $154$ & $168$ & $177$ & $140$ & $202$\\

$3.68$ & $62$ & $13$ & $8$ & $12$ & $17$ & $40$ & $23$ & $20$ & $17$ & $18$ & $14$ & $40$\\

$[1]$ & $[2]$ & $[3]$ & $[3]$ & $[3]$ & $[3]$ & $[2]$ & $[2]$ & $[3]$ & $[5]$ & $[5]$ & $[5]$ & $[5]$\\

\hline\hline
\end{tabular}
\end{center} 
\end{table*}

\subsection{Hubble Measurements}

The measurements of Hubble parameters can potentially to be a complementary probe in constraining cosmological parameters. The Hubble parameter characterizes the expansion rate of our universe at different redshifts, and depends on the differential age of the universe as a function of redshift:
\begin{equation}
H(z) = -\frac{1}{1+z}\frac{dz}{dt}~.
\end{equation}
Therefore, measuring the ${dz}/{dt}$ could straightforwardly estimate $H(z)$, which was firstly proposed by Ref. \cite{2002ApJ...573...37J}. They selected samples of passively evolving galaxies with high-quality spectroscopy, and then used stellar population models to constrain the age of the oldest stars in these galaxies. After that, they computed differential ages at different redshifts and obtained the determinations of Hubble parameter \cite{jimenez2003,simon2005}. Moreover, the Hubble parameter can also be obtained from the BAO measurement. By observing the typical acoustic scale in the light-of-sight direction, one can extract the expansion rate of the universe at certain redshift. Ref. \cite{gaztanaga2009} analyzed the information of Hubble parameter at redshift $z = 0.24$ and $z = 0.43$ from the Sloan Digital Sky Survey (SDSS) DR6 and DR7 data. Recently, these $H(z)$ data have been widely used on the determination of cosmological parameters, such as the effective number of neutrinos \cite{moresco2012,riemer2013}, the equation of state of dark energy \cite{lazkoz2007,pan2010,farooq2013}, the cosmography scenario \cite{xia2010,xia2012}, and the modified gravity models \cite{ben2011,zhangtj2012,aviles2013}.

In table \ref{Hz} we adopt 25 Hubble parameter data used in ref. \cite{zhengwei2014}. Furthermore, we also have the direct probe on the current Hubble constant $H_0$ obtained from the re-analysis of the ref. \cite{hstriess} Cepheid data made by ref. \cite{E2014} by using a revised geometric maser distance to NGC
4258 from \cite{ngc2013}: $H_0 = (70.6 \pm 3.3)~{\rm km~s^{-1}~Mpc^{-1}}$.

\subsection{BAO}

BAO provides an efficient method for measuring the expansion history by using features in the clustering of galaxies within large scale surveys as a ruler with which to measure the distance-redshift relation. Since the current BAO data are not accurate enough, one can only determine an effective distance
\begin{equation}
D_{\rm V}(z) = \left[(1+z)^2D_{\rm A}^2(z)cz/H(z)\right]^{1/3}~.
\end{equation}
In this paper, we use the BAO measurement of $r_{\rm drag}/D_{\rm V}(z)$, from the 6dF Galaxy Redshift Survey (6dFGRS) at a low redshift ($z = 0.106$) \cite{2011MNRAS.416.3017B}, the measurement $D_{\rm V}/r_{\rm drag}(z)$ of the BAO scale based on a re-analysis of the Luminous Red Galaxies (LRG) sample from SDSS Data Release 7 at the median redshift ($z = 0.35$) \cite{2010MNRAS.401.2148P}, the BAO signal of $D_{\rm A}(z)$ and $H(z)$ from BOSS CMASS DR9 data at redshift ($z = 0.57$) \cite{2014MNRAS.441...24A}, the BAO measurement of $D_{\rm V}/r_{\rm drag}(z)$ from the WiggleZ survey at $z = 0.44$, $z = 0.60$ and $z = 0.73$ \cite{2011MNRAS.418.1707B}, and the latest BAO measurement of $D_{\rm A}(z)$ and $H(z)$ at high redshift of $z = 2.34$ from the analysis of Ly-$\alpha$ forest of BOSS quasars \cite{2015A&A...574A..59D}.

\subsection{CMB Distance Information}

CMB measurement is sensitive to the distance to the decoupling epoch via the locations of peaks and troughs of the acoustic oscillations. Here we use the ``distance information'', following the Planck measurement \cite{planck2015fit}, which includes the ``shift parameter'' $R$, the ``acoustic scale'' $l_A$, and the photon decoupling epoch $z_\ast$. $R$ and $l_A$ correspond to the ratio of angular diameter distance to the decoupling era over the Hubble horizon and the sound horizon at decoupling, respectively, given by:
\begin{equation}
R=\frac{\sqrt{\Omega_mH^2_0}}{c}\chi(z_\ast)~,~~l_A=\frac{\pi\chi(z_\ast)}{\chi_s(z_\ast)}~,
\end{equation}
where $\chi(z_\ast)$ and $\chi_s(z_\ast)$ denote the comoving distance to $z_\ast$ and the comoving sound horizon at $z_\ast$, respectively. The decoupling epoch $z_\ast$ is given by ref. \cite{1996ApJ...471..542H}:
\begin{equation}
z_\ast=1048[1+0.00124(\Omega_b h^2)^{-0.738}][1+g_1(\Omega_m h^2)^{g_2}]~,
\end{equation}
where
\begin{equation}
g_1=\frac{0.0783(\Omega_b h^2)^{-0.238}}{1+39.5(\Omega_b
h^2)^{0.763}}~,~g_2=\frac{0.560}{1+21.1(\Omega_b h^2)^{1.81}}~.
\end{equation}
We calculate the likelihood of the CMB distance information as follows:
\begin{equation}
\chi^2=(x^{\rm th}_i-x^{\rm data}_i)(C^{-1})_{ij}(x^{\rm
th}_j-x^{\rm data}_j)~,
\end{equation}
where $x=(R,l_A,z_\ast)$ is the parameter vector and $(C^{-1})_{ij}$ is the inverse covariance matrix for the CMB distance information.

\section{Numerical Results}
\label{results}

In our analysis, we perform a global fitting using the COSMOMC package \cite{2002PhRvD..66j3511L}, a Monte Carlo Markov chain (MCMC) code. Besides the cosmic opacity parameter $\varepsilon$, we also vary the following cosmological parameters with top-hat priors: the dark matter energy density $\Omega_{\rm c}h^2$, the baryon energy density $\Omega_{\rm b}h^2$, the Hubble parameter $h$, and the constant dark energy equation of state $w$. For the JLA and SNLS datasets, we have two more nuisance parameters $\alpha$ and $\beta$ coming from light-curve calibration.

\subsection{Constant $\varepsilon$ from Various Datasets}

In the literature, the latest constraint on the cosmic opacity parameter $\varepsilon$ comes from the data combination of ``Union2 Compilation'' SNIa samples and the Hubble parameter as a function of $z$ (Hz) in ref. \cite{2010JCAP...10..024A}. It is worth to revisit the constraints on $\varepsilon$ from the recent measurements on SNIa and Hubble parameter, as well as some other useful probes, like GRBs, BAO and CMB distance information.

\begin{figure}[t]
\includegraphics[width=0.5\textwidth]{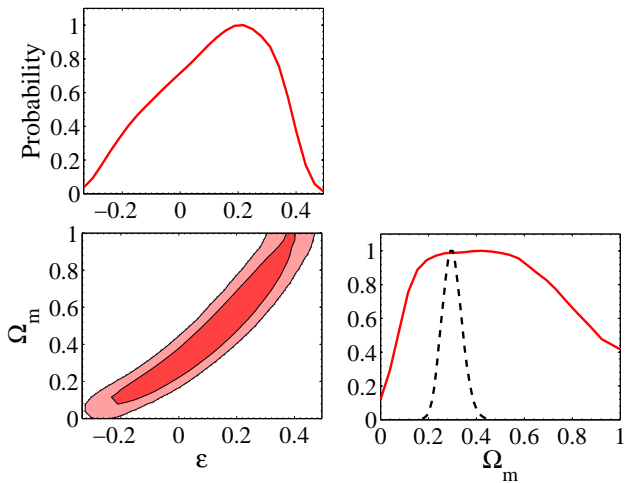}
   \caption{\label{eps-omm-1d-union}One and two dimensional constraints on the parameters $\varepsilon$ and $\Omega_{\rm m}$ from the Union2.1 SNIa data alone. For comparison, we also show the constraint on $\Omega_{\rm m}$ in the standard $\Lambda$CDM model with $\varepsilon=0$ (black dashed line).}
\end{figure}

Firstly, we use the latest SNIa datasets, Union2.1, JLA and SNLS Compilations, to obtain the limits on $\varepsilon$. In figure \ref{eps-omm-1d-union} we show the one-dimensional constraints on $\varepsilon$ and $\Omega_{\rm m}$ from Union2.1 data alone, as well as the two-dimensional contour between them. The constraint on $\varepsilon$ is very weak: $\varepsilon=0.11\pm 0.17$ (68\% C.L.), and seems not improved too much even using the latest SNIa sample, when comparing with the results in ref. \cite{2010JCAP...10..024A}. However, this constraint is slightly different from that result when using SNIa data alone. We also use the old Union2008 SNIa data alone to constrain opacity and obtain the similar result with ours. Therefore, this difference might be due to the different fitting methods, the different parameterizations or the different treatments on the systematics of Union data in the calculations.

In the two-dimensional contour of figure \ref{eps-omm-1d-union}, there is a strong positive correlation between $\varepsilon$ and $\Omega_{\rm m}$. The reason is that the larger value of $\varepsilon$ is, the more the flux received from the source is. Therefore, the supernovae are not so fainter than expected from a matter dominated Universe, which means the accelerating Universe might be not really necessary. When including the parameter $\varepsilon$, the constraint on $\Omega_{\rm m}$ is significantly enlarged, see the one-dimensional distribution of $\Omega_{\rm m}$ in figure \ref{eps-omm-1d-union}. If assuming the universe is transparent, the constraint on $\Omega_{\rm m}$ shrinks dramatically, $\Omega_{\rm m}=0.30\pm 0.04$ (68\% C.L.), from the Union2.1 data alone.

\begin{figure*}[t]
\includegraphics[width=0.33\textwidth]{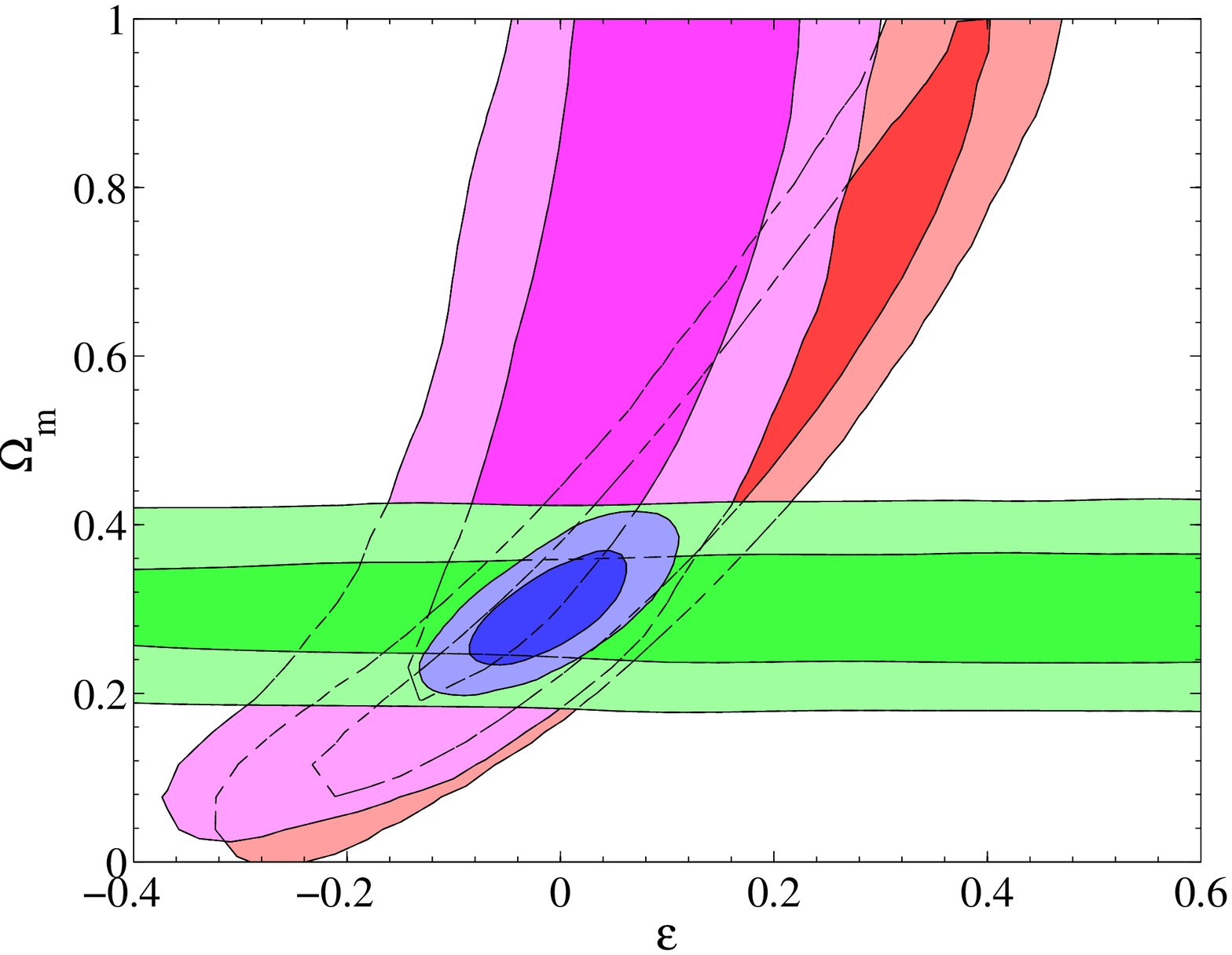}
\includegraphics[width=0.33\textwidth]{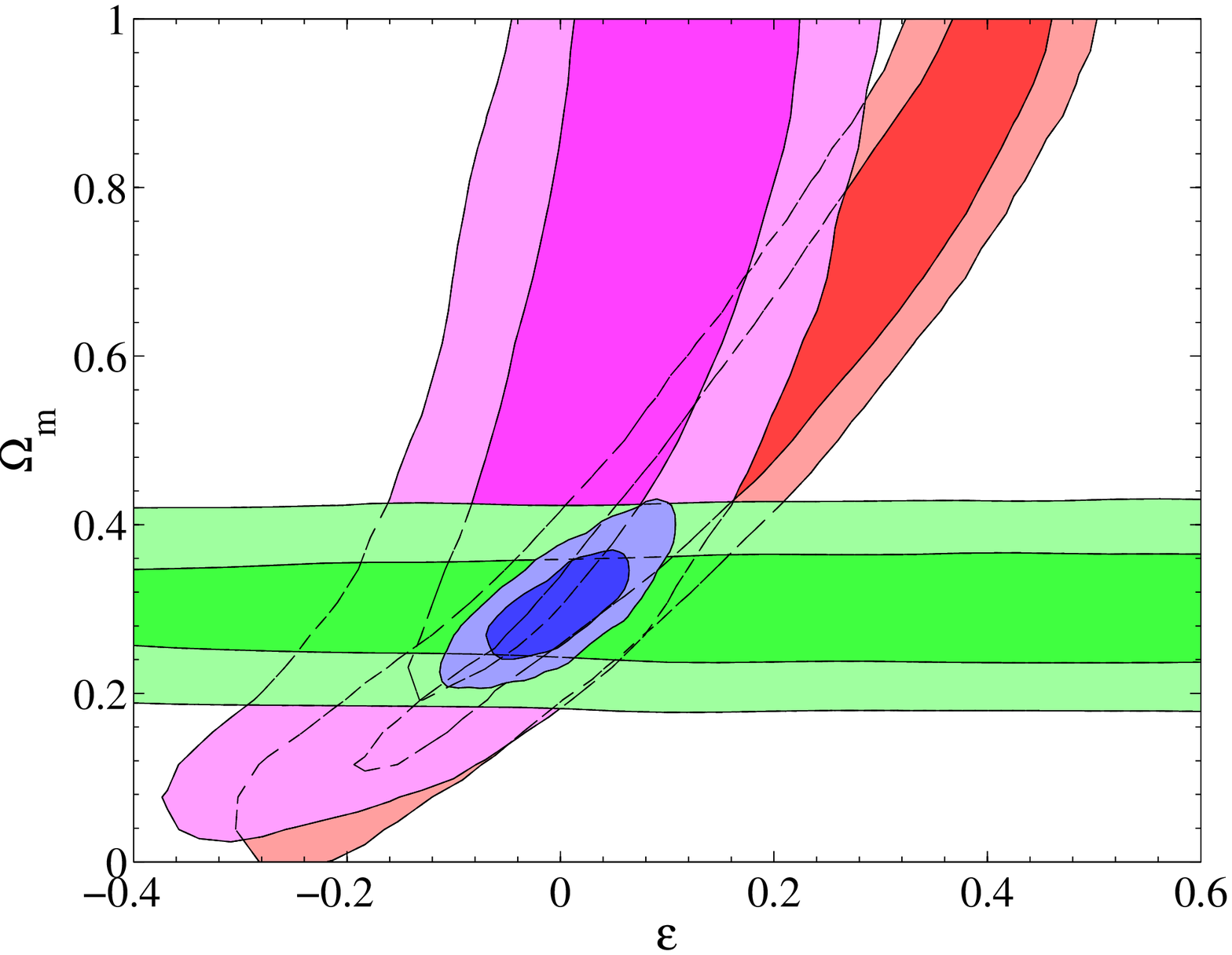}
\includegraphics[width=0.33\textwidth]{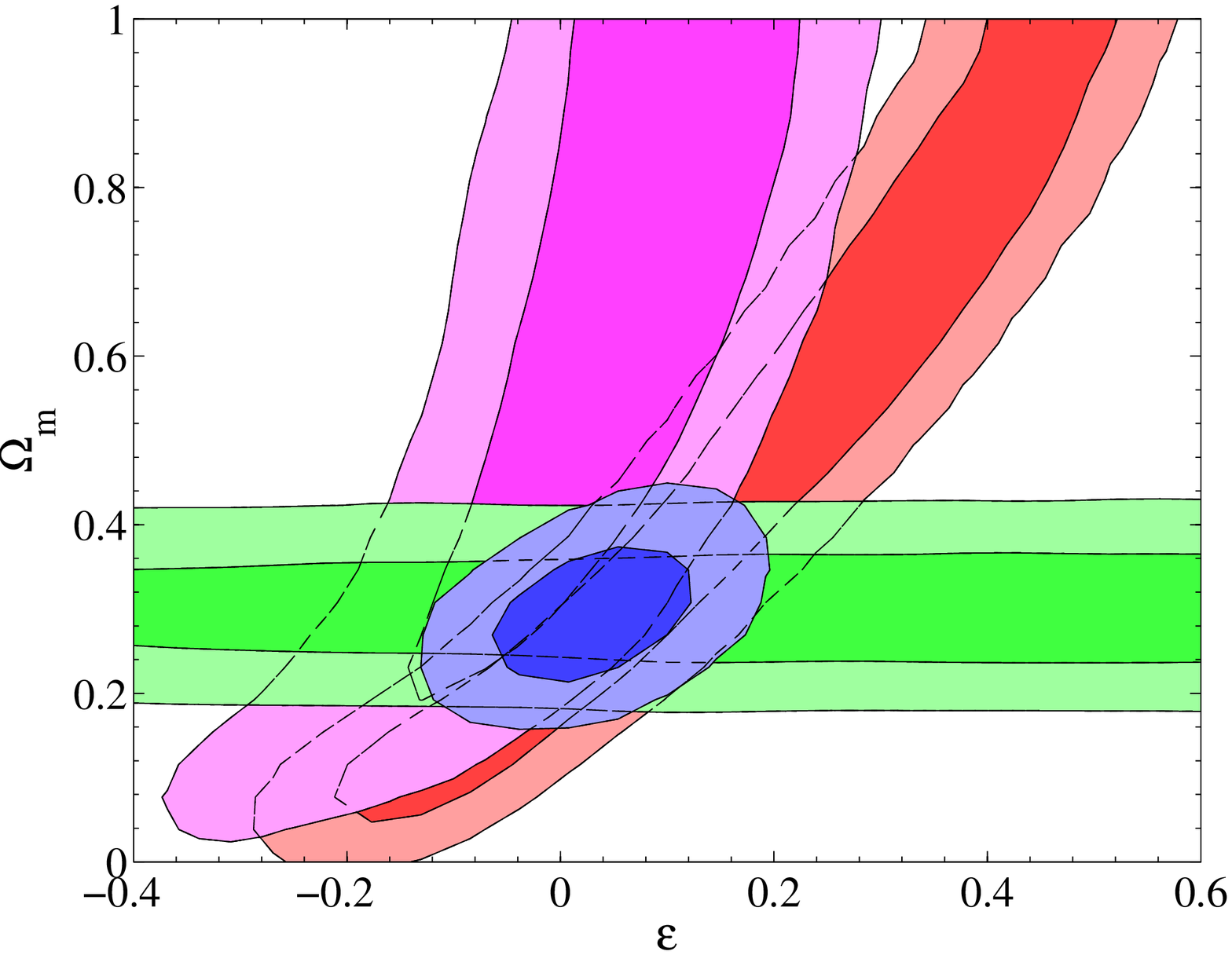}
   \caption{\label{eps-omm-2d-sn}Two-dimensional constraints on the panel ($\varepsilon$,$\Omega_{\rm m}$) from the Union2.1 (left panel), JLA (middle panel) and SNLS (right panel) data, respectively, as well as the GRBs and measurements of Hubble parameter.}
\end{figure*}

Besides the Union2.1 sample, there are two more SNIa samples, JLA and SNLS, which has been widely used in the recent literature. Therefore, we also use JLA and SNLS data to constrain the parameter $\varepsilon$, respectively. We obtain the similar constraints at 68\% confidence level: $\varepsilon=0.17\pm 0.18$ and $\varepsilon=0.19\pm 0.19$ from JLA and SNLS data. This result implies that the constraining power on $\varepsilon$ from the luminosity distance indicator provided by SNIa is hardly to be improved at present, due to the strong correlation with $\Omega_{\rm m}$. In figure \ref{eps-omm-2d-sn}, we show the two-dimensional contours between $\varepsilon$ and $\Omega_{\rm m}$ (red contours) from Union2.1, JLA and SNLS, respectively.

In order to improve the constraint on $\varepsilon$, we also include the GRB data into the calculations. GRBs is another kind of luminosity distance indicator, that can potentially be used to measure the luminosity distance out to higher redshift than SNIa. In figure \ref{eps-omm-2d-sn} we show the two-dimensional contour on the panel of ($\varepsilon$,$\Omega_{\rm m}$) from GRBs sample alone (magenta contours). Since it is also the luminosity distance indicator, the constraining power on $\varepsilon$ is limited as well, due to the strong correlation between $\varepsilon$ and $\Omega_{\rm m}$. However, due to its high redshift of GRB samples, the allowed region of large value of $\varepsilon$ is shrunk clearly: $\varepsilon=0.03\pm 0.13$ (68\% C.L.) and the significance of the models with large value of $\varepsilon$ is suppressed. Furthermore, the direction of the correlation between $\varepsilon$ and $\Omega_{\rm m}$ is also different from that obtained from SNIa data alone. When we use Union2.1 and GRB data together, the constraint on $\varepsilon$ is slightly improved: $\varepsilon=0.003\pm 0.12$ (68\% C.L.). Note that the GRBs sample suffers from the ``circular problem'' if using GRBs as cosmological standard ruler. We keep into account in our statistical analysis the issues related to the circular problem.

The Hubble parameter as a function of redshift (Hz) is an useful measurement which has been used in many works. Here, we use this dataset to study the cosmic opacity parameter $\varepsilon$. Since Hz data can give good constraint on $\Omega_{\rm m}$ and there is a strong positive correlation between $\varepsilon$ and $\Omega_{\rm m}$, Hz data can also significantly improves the constraint on $\varepsilon$ indirectly. In figure \ref{eps-omm-2d-sn} we show the constraint from Hz data (green contours), and apparently it has nothing to do with the constraint on $\varepsilon$ directly. However, if we combine the Hz data with SNIa or GRB datasets together, the constraint on $\varepsilon$ is dramatically shrunk, namely the 68\% C.L. constraint $\varepsilon=-0.008\pm 0.048$ from Union2.1, GRB and Hz data. This limit is similar with that obtained from Union2+Hz data in Ref. \cite{2010JCAP...10..024A}. We also show the constraints from JLA+GRB+Hz and SNLS+GRB+Hz data combinations in other two panels of figure \ref{eps-omm-2d-sn}, respectively. The constraints on $\varepsilon$ are quite similar, in which the JLA+GRB+Hz gives slightly better constraint, while SNLS+GRB+Hz gives slightly worse constraint.

Finally, we include the BAO and CMB distance information into the calculations to narrow the constraint on $\Omega_{\rm m}$ and consequently improve the limit of the cosmic opacity parameter $\varepsilon$. Firstly, we use the BAO measurements, together with the Union2.1, GRB and Hz data, to constrain $\varepsilon$ and find that the constraint on $\varepsilon$ is significantly improved: $\varepsilon=0.009\pm0.024$ ($1\sigma$). This constraint is tighter than that obtained in ref. \cite{More:2008uq}, due to several latest precise BAO measurements we use. Then we use the CMB distance information to replace the BAO measurements in the calculation. The CMB distance information contains messages from the early Universe at $z\sim1100$ which is clearly different from other probes at $z\lesssim 2.5$. But it can give tighter constraint on the matter energy density, and then indirectly affect the study on the cosmic opacity. Using the CMB distance information from Planck measurement, we obtain the even tighter constraint than that from BAO: $\varepsilon=0.025\pm0.020$ ($1\sigma$). When we combine all these data together, due to the strong constraining power on $\Omega_{\rm m}$ from BAO and CMB data, the constraint on $\varepsilon$ is significantly improved:
\begin{equation}
\varepsilon=0.023\pm 0.018~~~~(68\%~{\rm C.L.})~.\label{best}
\end{equation}
When comparing with the constraints in ref. \cite{2010JCAP...10..024A}, the statistical error bar of $\varepsilon$ has been shrunk by a factor of 2, due to the new BAO and CMB distance information in the calculations. In the meanwhile, since the new BAO and CMB data favor a large number of $\Omega_{\rm m}$ \cite{planck2015fit}, now the data combination slightly favors a positive value of $\varepsilon$, due to the positive correlation between $\varepsilon$ and $\Omega_{\rm m}$. But the significance level is only about 1.2$\sigma$.

\begin{figure}[t]
\includegraphics[width=0.5\textwidth]{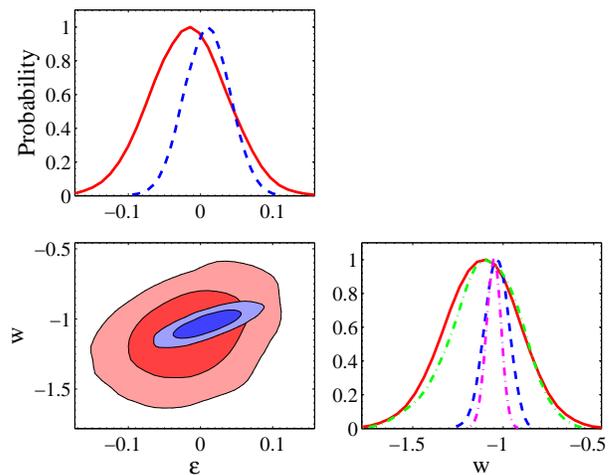}
   \caption{\label{eps-2d-corr}One and two dimensional constraints on the parameters $\varepsilon$ and $w$ from the Union2.1+GRB+Hz data (red) and Union2.1+GRB+Hz+BAO+ShR data (blue), respectively. For comparison, we also show the one-dimensional constraints on $w$ from these two data combinations in the standard $w$CDM model with $\varepsilon=0$.}
\end{figure}

More importantly, besides the correlation with $\Omega_{\rm m}$, $\varepsilon$ is also strongly correlated with the dark energy equation of state $w$. In figure \ref{eps-2d-corr}, we show the one-dimensional distributions of $w$ and $\varepsilon$ and two-dimensional contours between $w$ and $\varepsilon$ from the current data combinations. There is a positive correlation between $w$ and $\varepsilon$ from the Union2.1+GRB+Hz data combination. The reason is that the larger value of $\varepsilon$ is, the more the flux received from the source is. Then the supernovae are brighter than expected from the standard $\Lambda$CDM Universe with $w=-1$. Consequently, the large value of $w$ is favored by the data. The constraint on $\varepsilon$ is slightly weaker than the case discussed above, $\varepsilon=-0.016\pm0.053$ (68\% C.L.), due to the degeneracy. We also obtain the constraints on $w$, as shown in figure \ref{eps-2d-corr}: $w=-1.12\pm0.20$ and $w=-1.10\pm0.19$ for models with free $\varepsilon$ and with $\varepsilon=0$, respectively.

Furthermore, we include the more precise BAO and CMB data into the calculation and find that this positive correlation between $\varepsilon$ and $w$ becomes much stronger, as shown in the blue contours in figure \ref{eps-2d-corr}. When comparing with eq. \ref{best}, the limit on $\varepsilon$ becomes significantly weaker:
\begin{equation}
\varepsilon=0.009\pm0.031~~~~(68\%~{\rm C.L.})~.
\end{equation}
In the meanwhile, the constraints on $w$ are also quite different, namely at 68\% confidence level $w=-1.04\pm 0.07$ and $w=-1.05 \pm 0.04$ for the models with free $\varepsilon$ and with $\varepsilon=0$, respectively.

\begin{figure*}[t]
\includegraphics[width=0.33\textwidth]{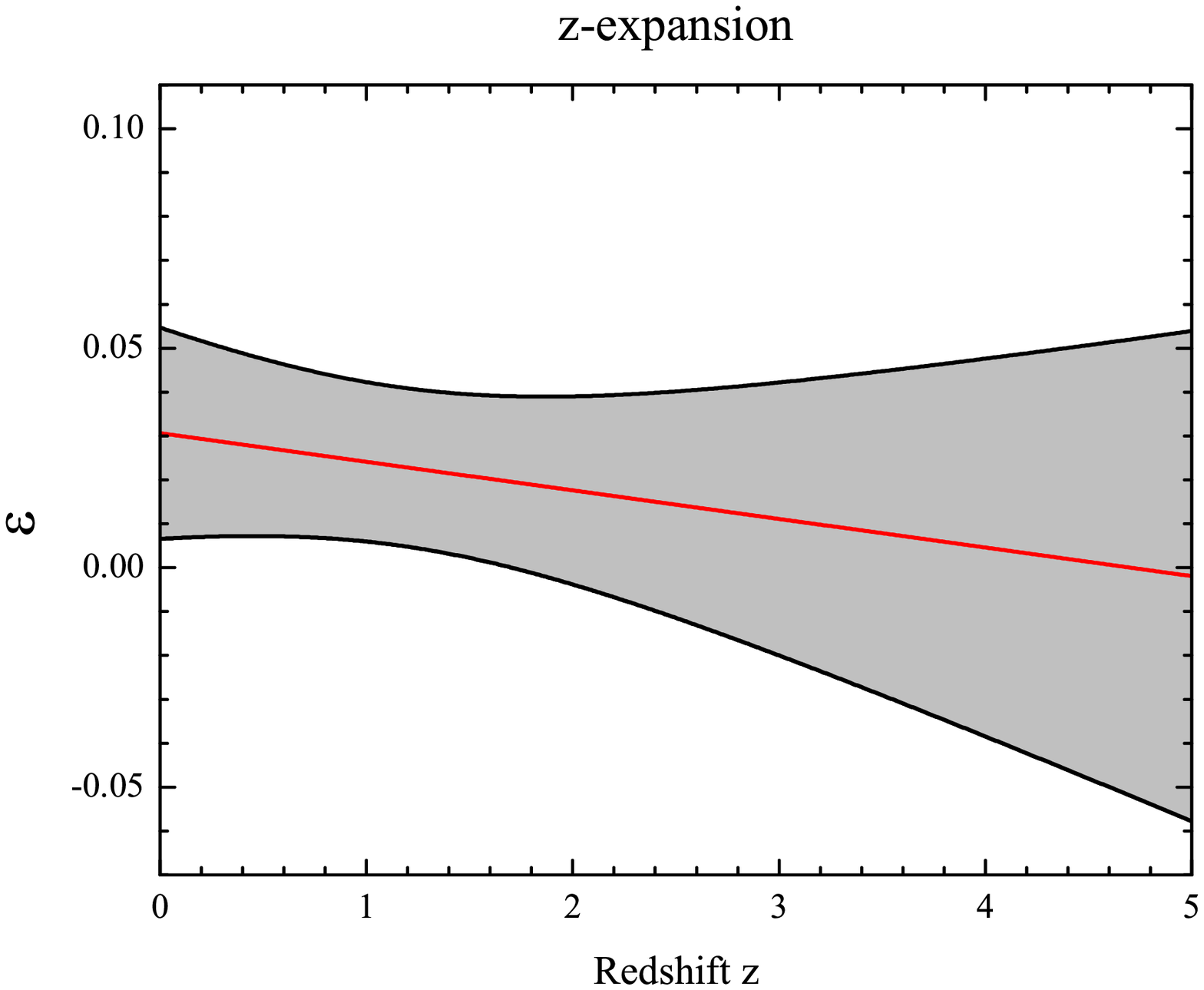}
\includegraphics[width=0.33\textwidth]{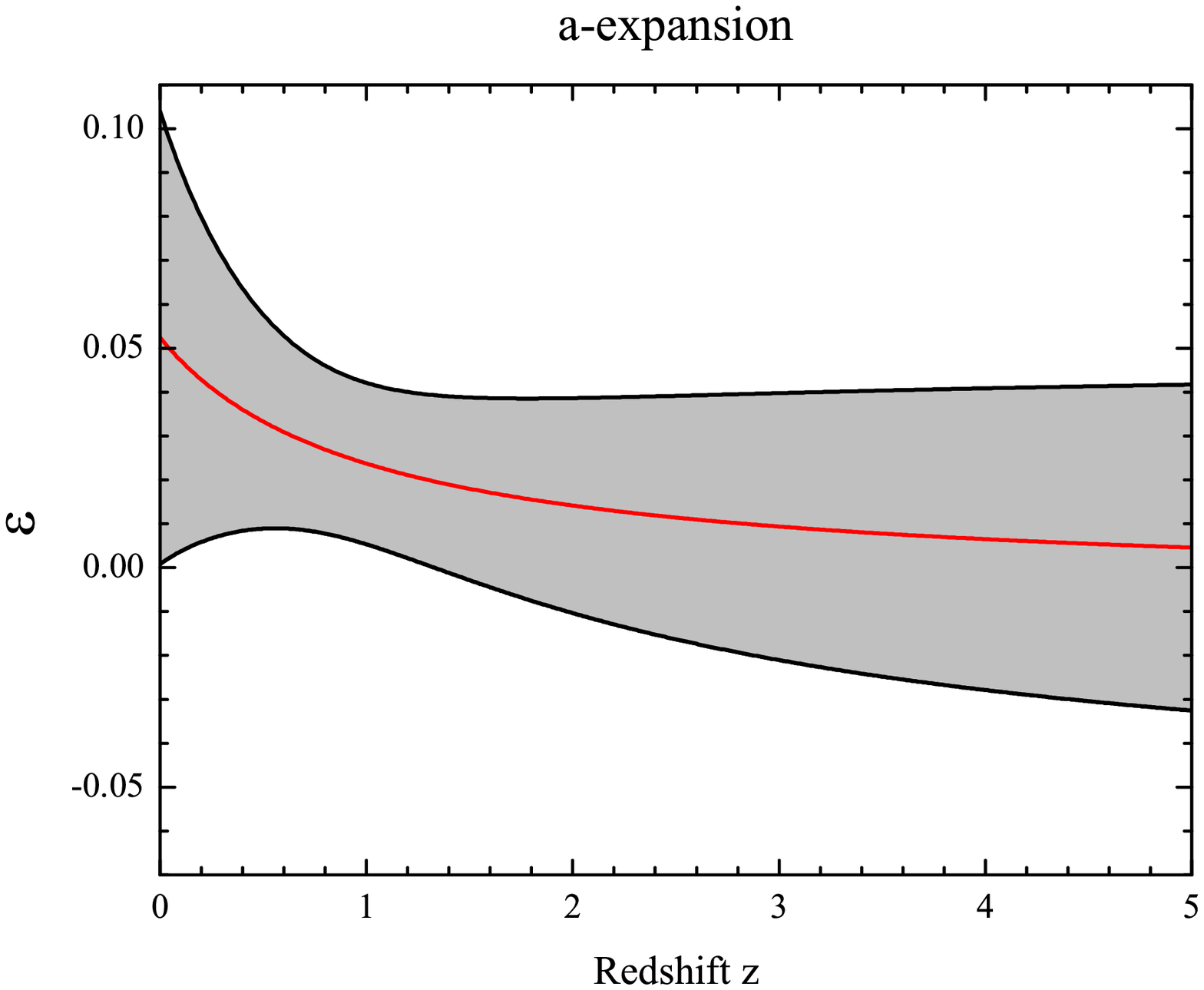}
\includegraphics[width=0.33\textwidth]{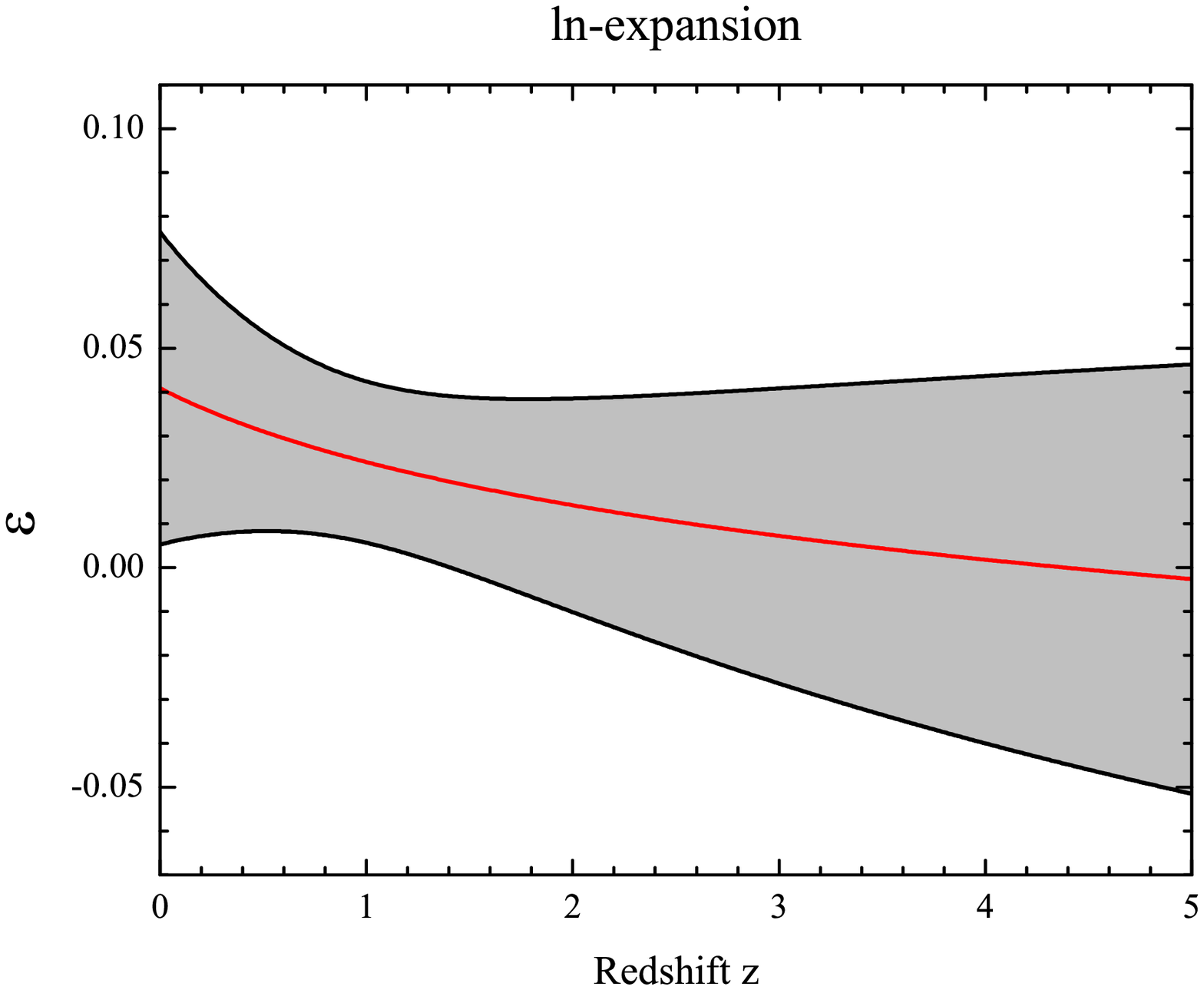}
   \caption{\label{para-all-exp}Constraints on the evolutions of cosmic opacity parameter $\varepsilon(z)$, median (central red line) and 68\% (dark shaded area) interval of $\varepsilon(z)$, using three parametrization forms.}
\end{figure*}

\subsection{Parametrization \& PCA}

In the above subsection, we only consider the constant cosmic opacity parameter $\varepsilon$. But generally speaking, $\varepsilon$ should vary as a function of redshift during the evolution of Universe. Therefore, in this subsection, we follow the parametrization idea in the dark energy studies and parameterize $\varepsilon$ in three forms \cite{2011RAA....11.1199C,Hees:2014lfa}:
\begin{itemize}
\item $\varepsilon = \varepsilon_{\rm z0} + \varepsilon_{\rm z}z$~,~~~~($z$-expansion);
\item $\varepsilon = \varepsilon_{\rm a0} + \varepsilon_{\rm a}(1-a)$~,~~~~($a$-expansion);
\item $\varepsilon = \varepsilon_{\rm ln0} + \varepsilon_{\rm ln}\ln(1+z)$~,~~~~($\ln$-expansion).
\end{itemize}
Using the Union2.1, GRB, Hz, BAO and CMB data together (all data), we obtain the constraints on the free parameters in three parametrization forms and list them in Table \ref{PARAresults}.

Firstly, we can see that the obtained median values on $\varepsilon_{\rm i0}$ in three parametrization forms are similar and all consistent with zero, namely the 68\% C.L. limits are $\varepsilon_{\rm z0}=0.030\pm0.024$, $\varepsilon_{\rm a0}=0.052\pm0.051$, and $\varepsilon_{\rm ln0}=0.041\pm0.036$ for three parametrizations, respectively. No matter which parametrization form we use, $\varepsilon_{\rm i0}$ always dentes the current value of cosmic opacity at $z=0$.

One the other hand, the constraints on $\varepsilon_{\rm i}$ for three parametrization forms are quite different for both the median values and the statistical error bars. When combining the constraints on $\varepsilon_{\rm i0}$ and $\varepsilon_{\rm i}$, we could plot the evolution behaviour of cosmic opacity $\varepsilon$ as a function of $z$, as well as the statistical error bars of $\Delta\varepsilon(z)$:
\begin{equation}
\Delta\varepsilon(z) = \sqrt{(\Delta\varepsilon_{\rm i0})^2 + 2\Delta\varepsilon_{\rm i0}\Delta\varepsilon_{\rm i}f_{\rm i}{\rm cov}(\varepsilon_{\rm i0},\varepsilon_{\rm i}) + (\Delta\varepsilon_{\rm i}f_{\rm i})^2}~,
\end{equation}
where $\Delta\varepsilon_{\rm i0}$ and $\Delta\varepsilon_{\rm i}$ denote the obtained $1\sigma$ statistical error bars on $\varepsilon_{\rm i0}$ and $\varepsilon_{\rm i}$, respectively, ${\rm cov}(\varepsilon_{\rm i0},\varepsilon_{\rm i})$ is the correlation between $\varepsilon_{\rm i0}$ and $\varepsilon_{\rm i}$, and $f_{\rm i}$ denotes $z$, $(1-a)$ and $\ln(1+z)$ for three parametrization forms, respectively. And we obtain the similar evolutions of $\varepsilon(z)$, which are consistent with zero, in the redshift region $z\in[0,5]$ for three parametrization forms, as shown in figure \ref{para-all-exp}.

\begin{table}[t] 
\caption{Current and Future constraints on the parameters in three parametrization forms of the cosmic opacity parameter $\varepsilon$.}\label{PARAresults}
\begin{center}
\begin{tabular}{c|c|c|c|c}

\hline
\hline

&\multicolumn{2}{c}{$\varepsilon_{\rm i0}$}&\multicolumn{2}{|c}{$\varepsilon_{\rm i}$}\\
\cline{2-5}
& ALL Data & Future Data & ALL Data & Future Data \\
\hline
$z$-exp. & $0.030\pm0.024$ & 0.018 & $-0.007\pm0.014$ & 0.011 \\
$a$-exp. & $0.052\pm0.051$ & 0.035 & $-0.057\pm0.097$ & 0.055 \\
$\ln$-exp. & $0.041\pm0.036$ & 0.025 & $-0.024\pm0.042$ & 0.026 \\
\hline\hline
\end{tabular}
\end{center} 
\end{table}

Since the current data are not accurate enough, it is worthwhile discussing the constraint on the cosmic opacity $\varepsilon$ from the future measurements. Here, we use the future SNIa experiment, WFIRST, and the BAO measurement in BOSS experiment (similar analysis using the Euclid measurement can be found in Refs. \cite{2010JCAP...10..024A,euclid}). The fiducial models are taken from the best-fit values by all data combination in the $\Lambda$CDM framework with $\varepsilon=0$.

According to the updated report by Science Definition Team \cite{2015arXiv150303757S}, we obtain 2725 SNIa over the region $0.1<z<1.7$ with a bin $\Delta z=0.1$ of the redshift. The photometric measurement error per supernova is $\sigma_{\rm meas} = 0.08$ magnitudes. The intrinsic dispersion in luminosity is assumed as $\sigma_{\rm int} = 0.09$ magnitudes (after correction/matching for light curve shape and spectral properties). The other contribution to statistical errors is gravitational lensing magnification, $\sigma_{\rm lens} = 0.07 \times z$ mags. The overall statistical error in each redshift bin is then
\begin{equation}
\sigma_{\rm stat} = \left[(\sigma_{\rm meas})^2 + (\sigma_{\rm int})^2 + (\sigma_{\rm lens})^2 \right]^{1/2} / \sqrt{N_i} ,
\end{equation}
where $N_i$ is the number of supernova in the $i$-th redshift bin. According to being estimated, a systematic error per bin is
\begin{equation}
\sigma_{\rm sys} = 0.01 (1+z) /1.8  .
\end{equation}
Therefore, the total error per redshift bin is
\begin{equation}
\sigma_{\rm tot} = \left[(\sigma_{\rm stat})^2 + (\sigma_{\rm sys})^2 \right]^{1/2} .
\end{equation}

\begin{figure*}[t]
\includegraphics[width=0.23\textwidth]{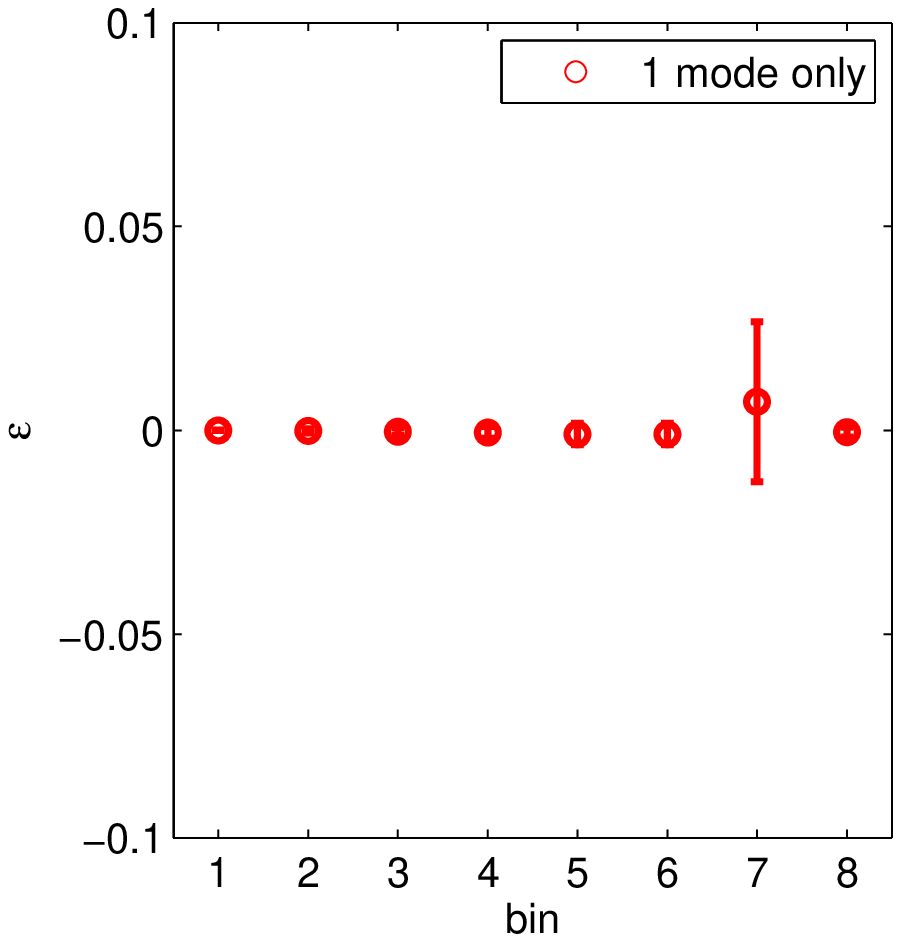}
\includegraphics[width=0.23\textwidth]{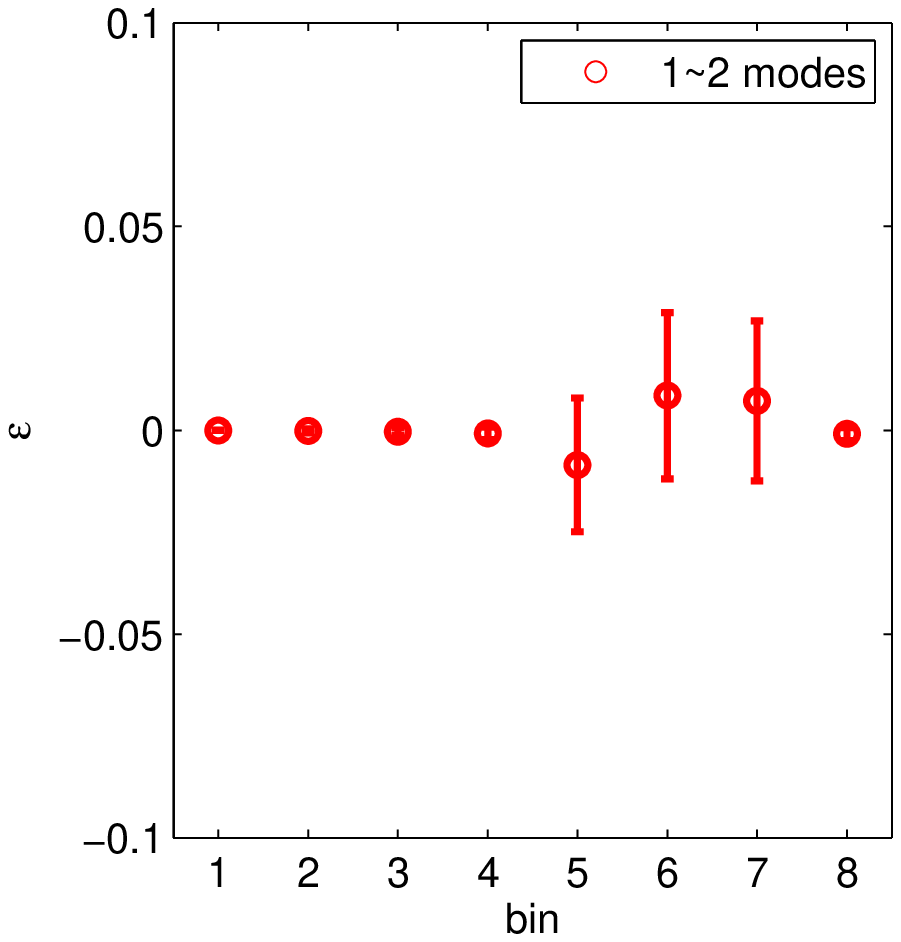}
\includegraphics[width=0.23\textwidth]{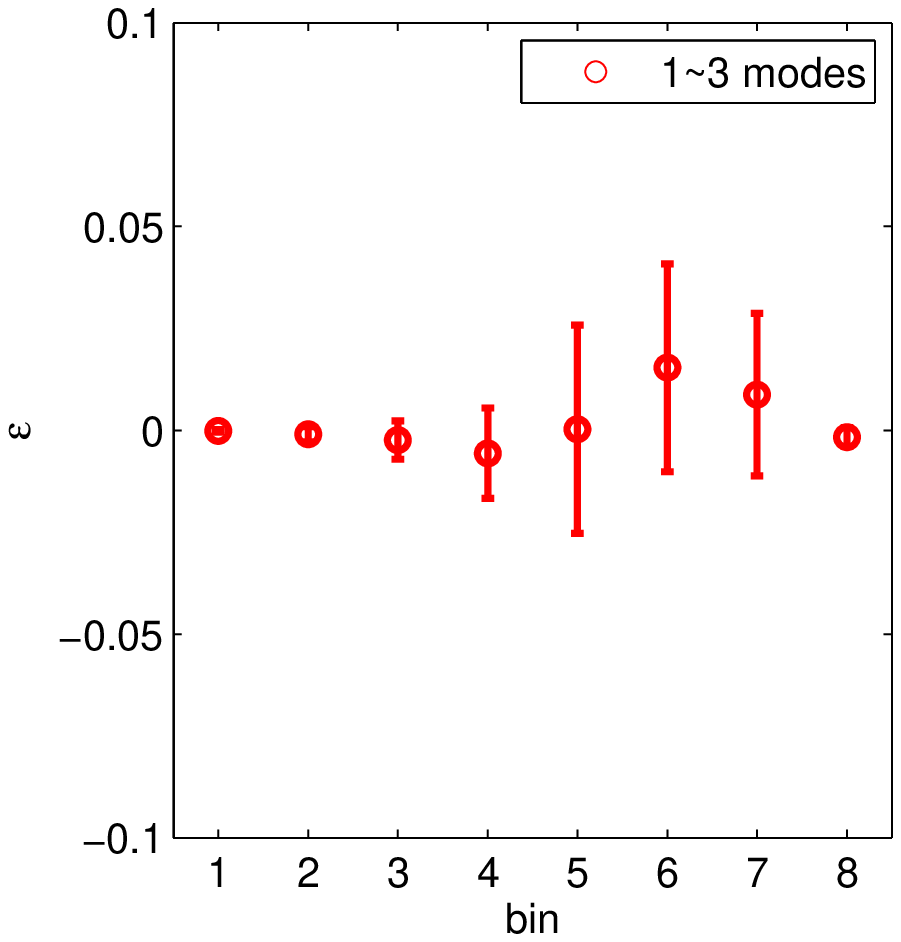}
\includegraphics[width=0.23\textwidth]{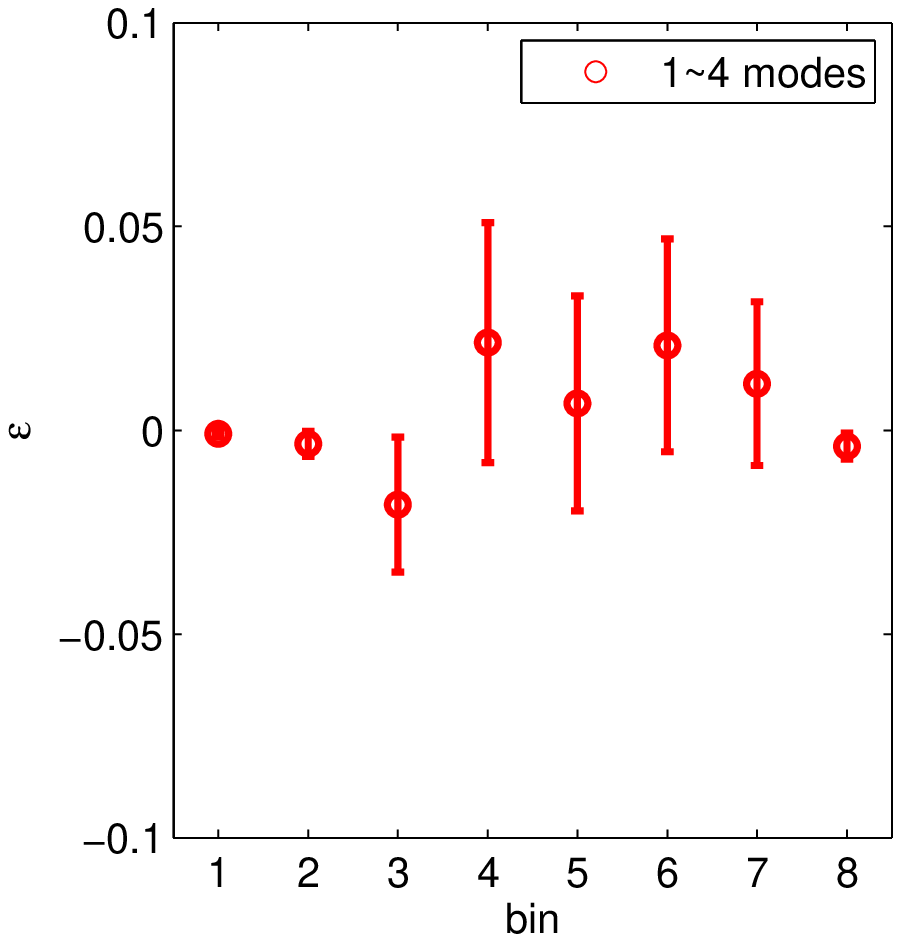}
\includegraphics[width=0.23\textwidth]{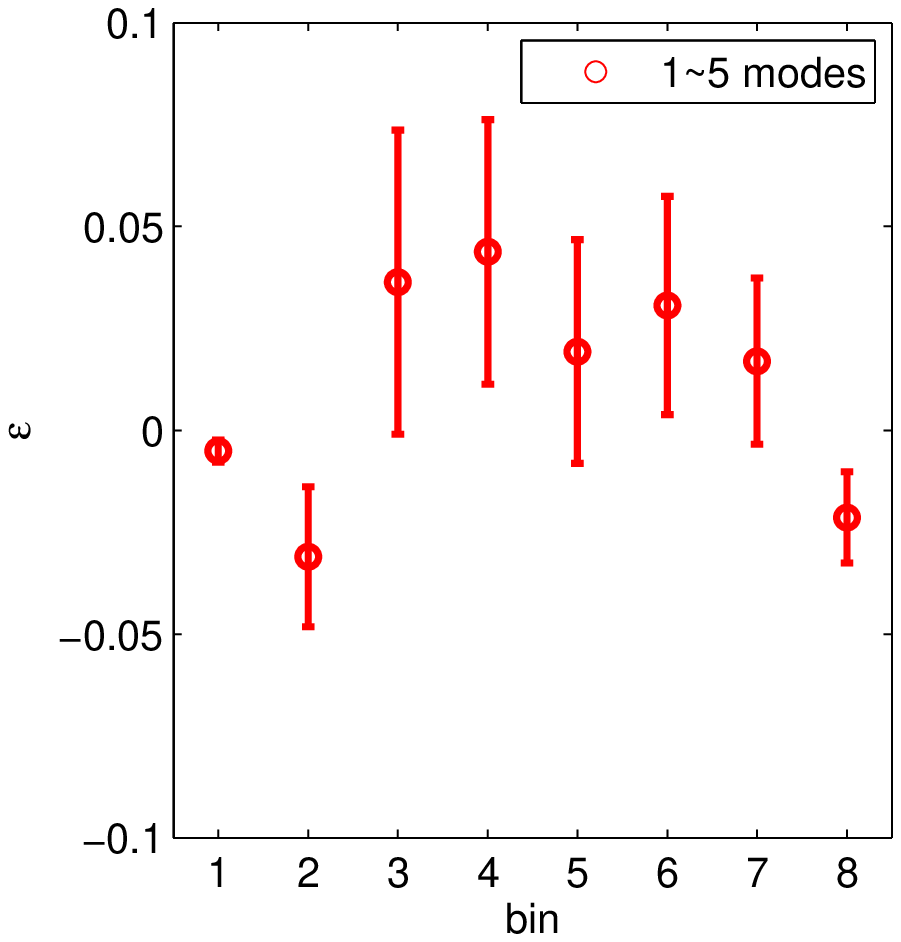}
\includegraphics[width=0.23\textwidth]{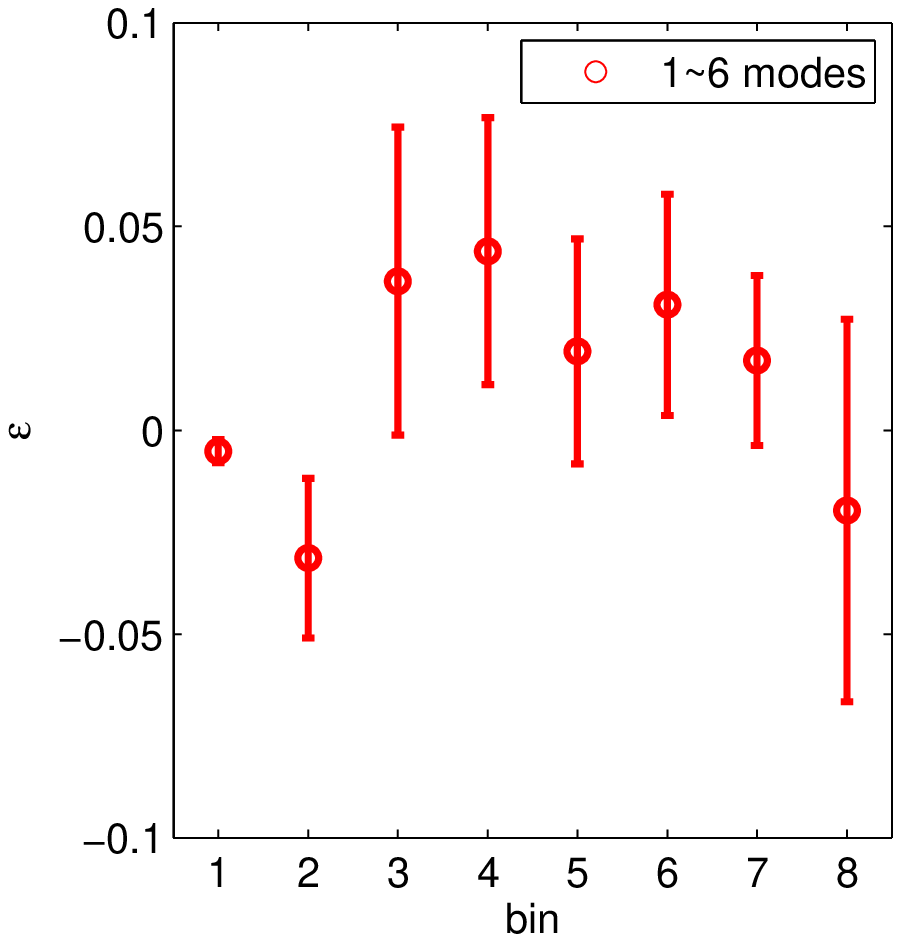}
\includegraphics[width=0.23\textwidth]{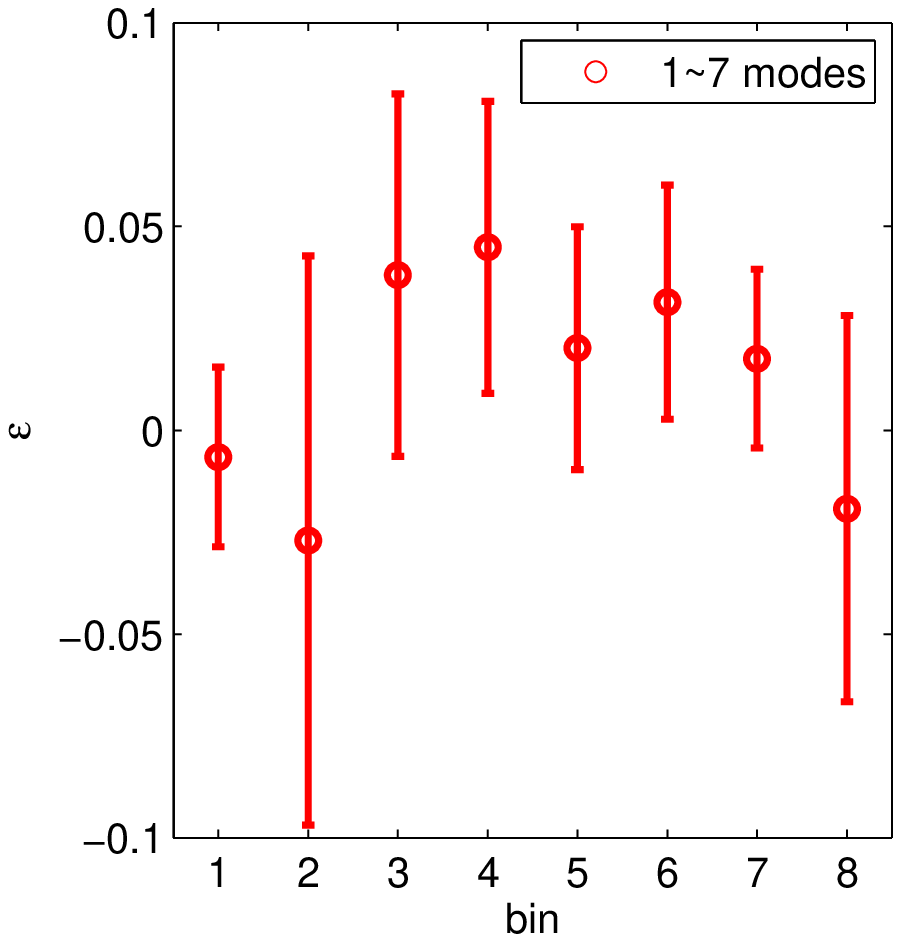}
\includegraphics[width=0.23\textwidth]{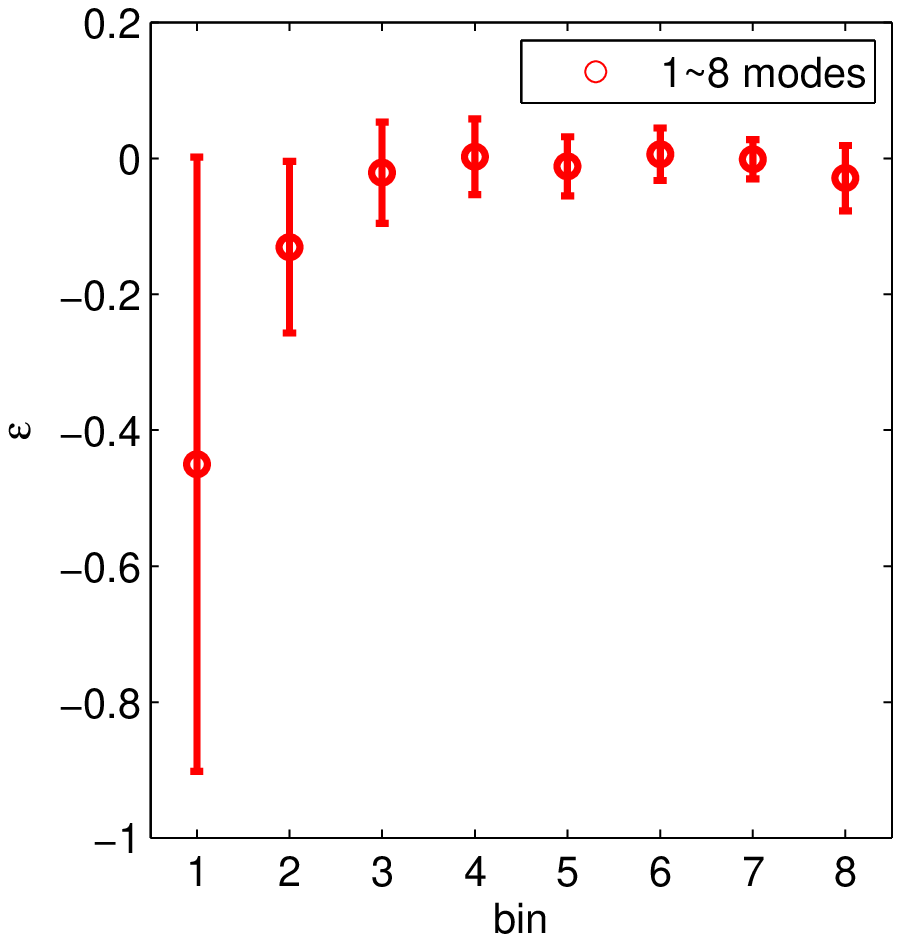}
   \caption{\label{pcamode}The reconstructed $\varepsilon(z)$ using the different number of eigenmodes.}
\end{figure*}

For the BAO simulation, we focus on the BOSS \cite{boss} which is part of the SDSS-III survey and is scheduled to operate over the period $2009-2014$. Using the $2.5$m SDSS telescope, it will measure redshifts of 1.5 million luminous galaxies in the range $0.1 < z < 0.7$ (as well as Ly¦Á absorption towards $160,000$ high-redshift quasars at about $z \sim 2.5$), covering $\sim10,000~{\rm deg}^2$ of high-latitude sky. The forecast precision for $H(z)$ is $1.8\%$, $1.7\%$ and $1.2\%$ in redshifts bins centered at $z = 0.35$, $0.6$ and $2.5$ respectively.

The standard deviation on the cosmic opacity parameter from the simulated mock data is $\Delta\varepsilon=0.005$ for the constant $\varepsilon$, corresponds to a transparency bound $\Delta\tau < 0.003$ (95\% C.L.) for the redshift between 0.2 and 0.35. This limit is six times tighter than the current result discussed in the above subsection. Besides the constant $\varepsilon$, we also constrain $\varepsilon(z)$ in three parametrization forms and list the standard deviations of $\varepsilon_{\rm i0}$ and $\varepsilon_{\rm i}$ in Table \ref{PARAresults}. Due to the improved precision of measurements, the future data could narrow the current constraints further by a factor of $\sim2$ and verify the evolution of the cosmic opacity parameter $\varepsilon(z)$.

Besides the parametrization method, in this paper we also perform a model-independent analysis through imposing several parameters $\varepsilon_n$ representing the cosmic opacity parameters in the different redshift bins.
Due to low precision of current observational data, the constraints on these coefficients are relatively weak.
In order to reduce dimension of parameter space, we adopt the principle component analysis (PCA) method \cite{Huterer2003,Huterer2005} which has been widely used in the cosmological data analyses.

In practice, we divide the redshift region of Union2.1 sample ($0<z<1.42$) into 7 bins and make sure there are similar number of SNIa samples in each bin. Therefore, we have 7 more free parameters $\varepsilon_n$ ($n=1,...,7$) to denote the cosmic opacity in each bin. Furthermore, we set another parameter $\varepsilon_8$ to denote the cosmic opacity at $z > 1.42$. Apparently, this parameter can only be constrained by the high redshift GRBs data. Using the all data combination, we obtain the constraint on $\varepsilon$ in each bin, as shown in the last plot of figure \ref{pcamode}. We find that the constraint on $\varepsilon_1$ is much weaker than those in other seven bins, which should be considered as the noise to be reduced by using the PCA method.

Firstly, we take an orthogonal transformation on original parameter space $\varepsilon_n$ to obtain a set of linearly uncorrelated variables $q_n$, reads:
\begin{eqnarray}\label{eqnpca}
F&=&W^{T}\Lambda W~,\nonumber\\
\mathbf{q}&=&W \mathbf{\varepsilon}~,
\end{eqnarray}
where $F$ is the fisher matrix describing the curvature of likelihood function in parameter space, $\Lambda$ is a diagonal matrix consisted of eigenvalues of $F$ and $W\in\mathrm{SO(8)}$ represents the transformation matrix. In practice, we obtain fisher matrix by inverting a covariance matrix generated from MCMC procedure, $F=C^{-1}$. Secondly, we reconstruct $\varepsilon(z)$ after truncating some noisy modes with small eigenvalues since large eigenvalue modes dominate variation of likelihood function. We find that most of eigenmodes have comparable eigenvalues, except the worst one which should be considered as noise effect.

Reconstructed $\varepsilon(z)$ with different number of modes are plotted in figure \ref{pcamode}. We can see that such large nonzero expected values in first two bins disappear when the last worst mode is dropped. For the reconstructed result, the less modes we take, the smaller influenced by noise we obtain. On the other hand, however, the less modes we take, the larger distortion between the PCA result and the real result we get. We should optimise how many modes we take in order to balance the amount of reduced noise and the lose of information that original result carries. Following the conventions of ref. \cite{Huterer2003}, it reads:
\begin{eqnarray}
risk &=& {bias}^2 + variance,\nonumber\\
     &=& \|\mathbf{\varepsilon}_{\rm{reconst}}-\mathbf{\varepsilon}\|_2^2 + \sum_{i=1}^{n}\sigma_i,\nonumber\\
     &=& \sum_{i=1}^{n}\left(\sum_{j=1}^8 W^T_{ij}q_j-\varepsilon_i\right)^2 + \sum_{i=1}^{n}\sigma_i
\end{eqnarray}
where ${bias}^2$ stands for quadratic reconstruction error, $\sigma_i$ is the diagonal element of matrix $\Lambda$ and $n$ is the number of used eigenmodes. In practice, we find that risk function reaches its minimum when $n=5$. Based on this result shown in figure \ref{pcamode}, there is no obvious deviation from zero in these reconstructions.

\section{Summary}
\label{summary}

The verification of the well-known Etherington relation is a useful way to search for the new physics beyond the standard model. Different from the model independent method, in which people use the information of $d_{\rm L}$ and $d_{\rm A}$ at the same redshift to constrain the parameter $\eta$, in this paper we adopt the model-dependent method, in which the photon number during the propagation is not conserved, and use the latest observational measurement, such as SNIa, GRB, Hz, BAO and CMB distance information, to study the cosmic opacity parameter $\varepsilon$. Here we summarize our main conclusions in more detail:

\begin{itemize}
\item By using the latest SNIa samples, such as Union2.1, JLA and SNLS, we find that the SNIa data alone can not constrain the cosmic opacity parameter $\varepsilon$ very well. The constraining power on $\varepsilon$ from the luminosity distance indicator provided by SNIa and GRB is hardly to be improved at present, due to the strong degeneracy between $\varepsilon$ and $\Omega_{\rm m}$.
\item The Hubble parameter as a function of redshift is an useful measurement which has been used in many works. Different from the SNIa and GRB data, Hz measurements can only indirectly improve the constraint on $\varepsilon$ using the strong correlation between $\varepsilon$ and $\Omega_{\rm m}$. When including the Hz data into the analysis, the statistical error bars are shrunk by a factor of $\sim 3$.
\item We also use the BAO and CMB distance information to study the constraint on $\varepsilon$ and obtain the tightest constraint: $\varepsilon=0.023\pm0.018$ (68\% C.L.). This constraint corresponds to a transparency bound $\Delta\tau < 0.018$ (95\% C.L.) for the redshift between 0.2 and 0.35.
\item There is a strong degeneracy between $\varepsilon$ and the equation of state of dark energy $w$. The larger value of $\varepsilon$ is, the more the flux received from the source is. Then the supernovae are brighter than expected from the standard $\Lambda$CDM Universe with $w=-1$.
\item Similar with the parametrization of dark energy equation of state, we also use three parametrization forms to denote the evolution of $\varepsilon$ as a function of $z$. Besides the parametrization, we also use the PCA method and find that there is no obvious deviation from zero in these reconstructions.
\item Finally, we simulate the future SNIa observation of WFIRST and the Hubble measurement in BOSS experiment. We obtain that the future mock data could give very tight constraint on the cosmic opacity $\varepsilon$ and verify the Etherington relation at high significance.
\end{itemize}

\acknowledgments We thank Yi-Fu Cai, Si-Yu Li, Taotao Qiu and Ming-Jian Zhang for useful discussions. J.-Q. Xia is supported by the National Youth Thousand Talents Program and the National Science Foundation of China under grant No. 11422323. The research is also supported by the Strategic Priority Research Program ``The Emergence of Cosmological Structures'' of the Chinese Academy of Sciences, grant No. XDB09000000.


\begin{thebibliography}{nn}

\bibitem{riess1998supernova}
A.~G.~Riess {\it et al.}, Astron. J. 116 (1998) 1009-1038.

\bibitem{perlmutter1999measurements}
S.~Perlmutter {\it et al.}, Astrophys.J. 517 (1999) 565-586.

\bibitem{planck2015fit}
P.~A.~R.~Ade {\it et al.}, [Planck Collaboration], arXiv:1502.01589 [astro-ph.CO].

\bibitem{Bassett:2003vu}
B.~A.~Bassett and M.~Kunz, Phys.\ Rev.\ D {\bf 69}, 101305 (2004).

\bibitem{Bassett:2003zw}
B.~A.~Bassett and M.~Kunz, Astrophys.\ J.\  {\bf 607}, 661 (2004).

\bibitem{Uzan:2004my}
J.~P.~Uzan, N.~Aghanim and Y.~Mellier, Phys.\ Rev.\ D {\bf 70}, 083533 (2004).

\bibitem{DeBernardis:2006ii}
F.~De Bernardis, E.~Giusarma and A.~Melchiorri, Int.\ J.\ Mod.\ Phys.\ D {\bf 15}, 759 (2006).

\bibitem{2010ApJ...722L.233H}
R.~F.~L.~Holanda, J.~A.~S.~Lima, M.~B.~Ribeiro, Astrophys. J. 722 (2010) L233-L237.

\bibitem{Nair:2011dp}
R.~Nair, S.~Jhingan and D.~Jain, JCAP {\bf 1105}, 023 (2011).

\bibitem{2011arXiv1104.2497L}
N.~Liang, Z.~Li, P.~Wu, S.~Cao, K.~Liao, Z.-H.~Zhu, Mon. Not. Roy. Astron. Soc. 436 (2013) 1017-1022.

\bibitem{2012ApJ...745...98M}
X.~L.~Meng, T.~J.~Zhang, H.~Zhan, Astrophys. J. 745 (2012) 98.

\bibitem{2011RAA....11.1199C}
S.~Cao, N.~Liang, Res. Astron. Astrophys. 11 (2011) 1199-1208.

\bibitem{2011SCPMA..54.2260C}
S.~Cao, Z.~H.~Zhu, Sci. China Phys. Mech. Astron. 54 (2011) 2260-2264.

\bibitem{Goncalves:2011ha}
R.~S.~Goncalves, R.~F.~L.~Holanda, J.~S.~Alcaniz, Mon. Not. Roy. Astron. Soc. 420 (2012) L43-L47.

\bibitem{Liao:2012bg}
K.~Liao, Z.~Li, J.~Ming, Z.~H.~Zhu, Phys. Lett. B 718 (2013) 1166-1170.

\bibitem{Ellis:2013cu}
G.~F.~R.~Ellis, R.~Poltis, J.~P.~Uzan ,A.~Weltman, Phys. Rev. D 87 (2013) 103530.

\bibitem{cddrzhu}
K.~Liao, Z.~Li, S.~Cao, M.~Biesiada, X.~Zheng, Z.-H.~Zhu, arXiv:1511.01318 [astro-ph.CO].

\bibitem{2014A&A...568A..22B}
M.~Betoule {\it et al.} [SDSS Collaboration], Astron. Astrophys. 568 (2014) A22.

\bibitem{More:2008uq}
S.~More, J.~Bovy, D.~W.~Hogg, Astrophys. J. 696 (2009) 1727-1732.

\bibitem{Nair:2012dc}
N.~Remya, J.~Sanjay, J.~Deepak, JCAP 1212 (2012) 028.

\bibitem{cddrpara}
A.~Avgoustidis, L.~Verde, R.~Jimenez, JCAP 0906 (2009) 012.

\bibitem{2010JCAP...10..024A}
A.~Avgoustidis, C.~Burrage, J.~Redondo, L.~Verde, R.~Jimenez, JCAP 1010 (2010) 024.

\bibitem{2012ApJ...746...85S}
N.~Suzuki {\it et al.}, Astrophys. J. 746 (2012) 85.

\bibitem{2011ApJS..192....1C}
A.~Conley {\it et al.} [SNLS Collaboration], Astrophys. J. Suppl. 192 (2011) 1.

\bibitem{2004ApJ...613L..13G}
G.~Ghirlanda, G.~Ghisellini, D.~Lazzati, C.~Firmani, Astrophys. J. 613 (2004) L13-L16.

\bibitem{2008ApJ...680...92L}
H.~Li, J.~Q.~Xia, J.~Liu, G.~B.~Zhao, Z.~H.~Fan, X.~Zhang, Astrophys. J. 680 (2008) 92-99.

\bibitem{firmani2007}
C.~Firmani, V.~Avila-Reese, G.~Ghisellini, G.~Ghirlanda, Rev. Mex. Astron. Astrofis. 43 (2007) 203.

\bibitem{schaefer2007}
B.~E.~Schaefer, Astrophys. J. 660 (2007) 16.

\bibitem{2001A&A...380....6G}
M.~Goliath, R.~Amanullah, P.~Astier, A.~Goobar, R.~Pain, Astron. Astrophys. 380 (2001) 6-18.

\bibitem{pietro2003}
E.~Di~Pietro, J.~F.~Claeskens, Mon. Not. R. Astron. Soc. 341 (2003) 1299.

\bibitem{2002ApJ...573...37J}
R.~Jimenez, A.~Loeb, Astrophys. J. 573 (2002) 37-42.

\bibitem{jimenez2003}
R.~Jimenez, L.~Verde, T.~Treu, D.~Stern, Astrophys. J. 593 (2003) 622.

\bibitem{simon2005}
J.~Simon, L.~Verde, R.~Jimenez, Phys. Rev. D 71 (2005) 123001.

\bibitem{gaztanaga2009}
E.~Gaztanaga, A.~Cabre, L.~Hui, Mon. Not. R. Astron. Soc. 399 (2009) 1663.

\bibitem{moresco2012}
M.~Moresco, L.~Verde, L.~Pozzetti, R.~Jimenez, A.~Cimatti, J. Cosmol. Astropart. Phys. 1207 (2012) 053.

\bibitem{riemer2013}
S.~Riemer-Sorensen, D.~Parkinson, T.~M.~Davis, C.~Blake, Astrophys. J. 763 (2013) 89.

\bibitem{lazkoz2007}
R.~Lazkoz, E.~Majerotto, J. Cosmol. Astropart. Phys. 0707 (2007) 015.

\bibitem{pan2010}
N.~Pan, Y.~Gong, Y.~Chen, Z.-H.~Zhu, Class. Quantum Grav. 27 (2010) 155015.

\bibitem{farooq2013}
O.~Farooq, D.~Mania, B.~Ratra, Astrophys. J. 764 (2013) 138.

\bibitem{xia2010}
V.~Vitagliano, J.-Q.~Xia, S.~Liberati, M.~Viel, J. Cosmol. Astropart. Phys. 1003 (2010) 005.

\bibitem{xia2012}
J.-Q.~Xia, V.~Vitagliano, S.~Liberati, M.~Viel, Phys. Rev. D 85 (2012) 043520.

\bibitem{ben2011}
G.~R.~Bengochea, Phys. Lett. B 695 (2011) 405.

\bibitem{zhangtj2012}
H.~Wang, T.-J.~Zhang, Astrophys. J. 748 (2012) 111.

\bibitem{aviles2013}
A.~Aviles, A.~Bravetti, S.~Capozziello, O.~Luongo, Phys. Rev. D 87 (2013) 044012.

\bibitem{zhengwei2014}
W.~Zheng, H.~Li, J.-Q.~Xia, Y.-P.~Wan, S.-Y.~Li, Int. J. Mod. Phys. D 23 (2014) 1450051.

\bibitem{hstriess}
A.~G.~Riess {\it et al.}, Astrophys. J. 730 (2011) 119.

\bibitem{E2014}
G.~Efstathiou, Mon. Not. Roy. Astron. Soc. 440 (2014) 1138-1152.

\bibitem{ngc2013}
E.~M.~L.~Humphreys, M.~J.~Reid, J.~M.~Moran, L.~J.~Greenhill, A.~L.~Argon, Astrophys. J. 775 (2013) 13.

\bibitem{2011MNRAS.416.3017B}
F.~Beutler {\it et al.}, Mon. Not. Roy. Astron. Soc. 416 (2011) 3017-3032.

\bibitem{2010MNRAS.401.2148P}
W.~J.~Percival {\it et al.} [SDSS Collaboration], Mon. Not. Roy. Astron. Soc. 401 (2010) 2148-2168.

\bibitem{2014MNRAS.441...24A}
L.~Anderson {\it et al.} [BOSS Collaboration], Mon. Not. Roy. Astron. Soc. 441 (2014) 24-62.

\bibitem{2011MNRAS.418.1707B}
C.~Blake {\it et al.}, Mon. Not. Roy. Astron. Soc. 418 (2011) 1707-1724.

\bibitem{2015A&A...574A..59D}
T.~Delubac {\it et al.} [BOSS Collaboration], Astron. Astrophys. 574 (2015) A59.

\bibitem{1996ApJ...471..542H}
W.~Hu, N.~Sugiyama, Astrophys.J. 471 (1996) 542-570.

\bibitem{2002PhRvD..66j3511L}
A.~Lewis, S.~Bridle, Phys.Rev. D66 (2002) 103511.

\bibitem{Hees:2014lfa}
A.~Hees, O.~Minazzoli and J.~Larena, Phys.\ Rev.\ D {\bf 90}, 124064 (2014).

\bibitem{euclid}
L.~Amendola {\it et al.} [Euclid Theory Working Group Collaboration], Living Rev.\ Rel.\  {\bf 16}, 6 (2013).

\bibitem{2015arXiv150303757S}
D.~Spergel {\it et al.}, arXiv:1503.03757 [astro-ph.IM].

\bibitem{boss}
D.~Schlegel {\it et al.} [with input from the SDSS-III Collaboration], arXiv:0902.4680 [astro-ph.CO].

\bibitem{Huterer2003}
D.~Huterer, G.~Starkman, Phys. Rev. Lett. 90 (2003) 031301.

\bibitem{Huterer2005}
D.~Huterer, A.~Cooray, Phys. Rev. D 71 (2005) 023506.

\end{thebibliography}
\end{document}